# Model of single-electron performance of micro-pixel avalanche photo-diodes

Z. Sadygov[a], Kh. Abdullaev[a], G. Akhmedov[a], F. Akhmedov[a], S. Khorev[b,*],
R. Mukhtarov[a], A. Sadigov[a], A. Sidelev[a], A. Titov[a], F. Zerrouk[b], and V. Zhezher[a]

[a] *Joint Institute for Nuclear Research,
Dubna, Moscow Region, Russia*
[b] *Zecotek Photonics, Inc.
Richmond, BC, Canada*
*E-mail*: SKhorev@zecotek.com

ABSTRACT: An iterative model of the avalanche process in a micro-pixel avalanche photo-diode initiated by a single photo-electron is presented. The model describes development of the avalanche process in time, taking into account change of electric field within the depleted region caused by internal discharge and external recharge currents. Conclusions obtained as a result of modelling are compared with experimental data. Simulations show that typical durations of the front and rear edges of the discharge current have the same magnitude of less than 50 ps. The front of the external recharge current has the same duration; however duration of the rear edge depends on value of the quenching micro-resistor. It was found that effective capacitance of the pixel calculated as the slope of linear dependence of the pulse charge on bias voltage exceeds its real capacitance by a factor of two, while the total pixel voltage drop equals twice the value of bias over-voltage.

---

[*] Corresponding author.

## Contents

## 1. Introduction

Silicon micro-pixel avalanche photo-diodes (MAPD) also known as Geiger-mode avalanche photo-diodes (GM APD), micro-pixel photon counters (MPPC), and silicon photomultipliers (SiPMs) are widely used as photo-detectors both in scientific and industrial applications [1]–[5]. Their general design proposed in [6] includes a matrix of small *p-n* junction areas (pixels) with typical size of 10 to 100 μm created on the surface of a silicon substrate. These pixels are separated from each other by a certain space in order to eliminate charge coupling between them and each is connected to a common conductor through its individual micro-resistor with resistance of $10^5$ to $10^6$ Ω. The pixel area and the value of its micro-resistor are chosen so that the probability of spontaneous (dark) generation of charges in its active area is sufficiently small (≪ 1) over the time of electric relaxation of the pixel capacitance. This allows such devices to operate in over-voltage conditions, *i.e.* at reverse bias exceeding the breakdown value.

A great number of experimental and theoretical works (*e.g.* [7]–[16]) studied avalanche photo-diode behaviour in over-voltage conditions. Refs. [7]–[10] explored mechanisms of formation of micro-plasma current pulses with a typical flat top amplitude and random duration. The most comprehensive explanation of the physical origin of micro-plasma pulses was given by R. H. Haitz in [8].

R. H. Haitz's model includes voltage source $U_{\text{br}}$ (photo-diode breakdown potential), resistance $R_s$ (resistance of the space charge region of the photo-diode), and bi-stable switch $S$ connected in series. Internal photo-diode capacity $C_p$ is connected in parallel to these components. In order to observe micro-plasma pulses, another voltage source with voltage $U_d > U_{\text{br}}$ was connected to this circuit through a ballast resistor $R_p \leq R_s$. In this case, current pulses with a flat top and random duration emerge in the external circuit of the photo-diode. The external charge current value $J$ within the flat top of these pulses is given by expression $J = (U_d - U_{\text{br}})/(R_p + R_s)$, while the potential difference $U_p$ between the photo-diode terminals is equal to $U_p = (U_d - J \times R_p) \geq U_{\text{br}}$. This means that a self-sustained avalanche process takes place inside the modelled photo-diode with equal currents discharging ($I$) and charging ($J$) the capacitance $C_p$. This mode of operation happens at relatively large values of current $J$ (within the range of 50–100 μA depending on the device design) provided by voltage source $U_d$ and ballast resistor $R_p$. In his follow-up article [9], R. H. Haitz further demonstrated experimentally that at sufficiently large values of ballast resistance $R_p$, it is possible to observe short photo-signals of relatively constant amplitude and duration, but without any flat top. This mode of operation is usually referred to as Geiger mode. However, the author did not propose a corresponding model describing this mode of photo-diode operation.

Single-element avalanche photo-diodes operating in Geiger mode, also known as single-photon avalanche photo-diodes (SPAD), have been thoroughly studied in [11]–[16] for photon-counting applications. The authors of these publications introduced several new models (equivalent electrical circuits) of SPAD counters based on the original model proposed by R. H. Haitz. However, as it was mentioned above, Haitz's model cannot describe the Geiger-mode photo-diode because large values of resistance $R_p$ ($R_p \gg R_s$) do not allow the device to reach the conditions of micro-plasma breakdown. Therefore, to provide a valid description of the avalanche parameters, a photo-diode model needs to take into account development of the avalanche process occurring within the depleted region of the device at small recharge current values.

Refs. [17], [18] discuss results of numerical modelling of SPAD parameters, but no model of the avalanche process is provided. Nevertheless, the authors claim good agreement of their experimental data with their results of modelling of the device photo-response. Another model of avalanche process proposed in [19] is incorrect since it gave a value of critical load resistance required for operation in Geiger mode that is two orders of magnitude smaller than that of experimental samples. This problem was noticed by authors themselves in their following paper [20]. They hypothesised that an avalanche process initiated by a single photo-electron may be localised in a small (approximately 1×1 μm$^2$) part of the pixel area and, therefore, surface resistance of the pixel cathode should be taken into account. This suggestion is rebutted by experiments showing that the avalanche process is actually spread across nearly entire pixel area, and therefore gain of a single electron in MAPD increases proportionally with a pixel area [1], [3].

The present work proposes a new model of operation of devices such as SPAD and MAPD and compares the generated theoretical parameters with experimental data.

## 2. Model of avalanche process in micro-pixel avalanche photo-diode

Micro-pixel avalanche photo-diode consists of an array of identical *p-n* junctions (pixels). Each pixel acting as an independent photo-diode is connected to a common bias via its individual micro-resistor. Because of this, it is sufficient to consider operation of a single pixel for modelling of the multi-pixel device parameters. To simplify the following discussion, we will assume that the MAPD pixel has a $p^+$-*i*-$n^+$ structure with its individual quenching micro-resistor $R_p$ (**Figure 1**). Reverse bias voltage $U_d$ applied to the pixel creates in the *i*-layer of thickness $W$ a uniform electric field $E = E_d = U_d/W$ strong enough to start the avalanche process.

A single photo-electron created within the *i*-layer near the pixel's anode ($p^+$-layer) creates electron-hole pairs on its way through entire thickness ($W$) of the *i*-layer. Due to the exponential character of avalanche process, the majority of these electron-hole pairs are created near the cathode ($n^+$-layer, **Figure 1***a*). The same behaviour is followed in avalanche processes initiated by a single hole. In this latter case, the majority of electron-hole pairs are created near the pixel anode ($p^+$-layer). This circumstance leads us to make a model assumption that the impact ionisation process takes place only in thin layers (thickness $d \ll W$) near the cathode and anode of the pixel. Gain factor for a single electron and a single hole was calculated respectively as $M_e = \exp(\alpha W)$ and $M_h = \exp(\beta W)$. Here $\alpha$ and $\beta$ are ionisation coefficients for electrons and holes. The following expressions were used to take into account dependence of $\alpha$ and $\beta$ on electric field $E$ [21]:

$$\alpha(E) = 3.8 \times 10^6 \times \exp\left(-\frac{1.75 \times 10^6}{E}\right),$$
$$\beta(E) = 2.25 \times 10^7 \times \exp\left(-\frac{3.26 \times 10^6}{E}\right),$$
(1)

where electric field $E$ is in [V/cm] and $\alpha$ and $\beta$ are in [1/cm].

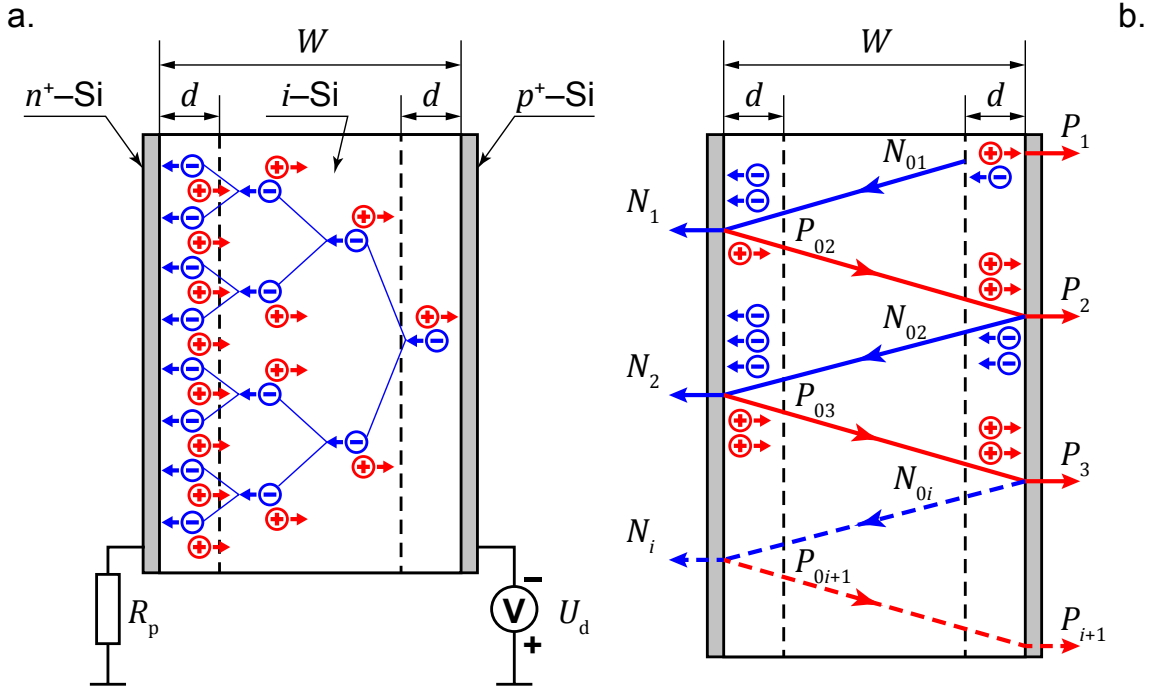

**Figure 1.** Schematic model of operation of a single MAPD pixel.

Successive stages of the modelled avalanche process are shown schematically in **Figure 1b**. One photo-electron is created near the anode ($N_{01} = 1$) at time $t = 0$. The electron passes through the *i*-layer and creates electron-hole pairs near the cathode. The time this process takes is $\tau = (W/v)$, where $v$ is drift velocity which is equal to the thermal velocity of charge carriers in the *i*-layer. Here, we have taken into consideration the fact that high electric fields effectively lead to saturation of drift velocities of both electrons and holes when they reach their maximal value, in silicon equal to $v \sim 10^7$ cm/s at room temperature. The number of electrons collected at the cathode $N_1$, number of holes moving toward the anode $P_{02}$, and electric field $E_1$ during this step can be expressed as

$$N_1 = \exp(\alpha_1 W), \qquad P_{02} = [\exp(\alpha_1 W) - 1], \qquad E_1 = (U_1/W). \tag{2}$$

where $\alpha_1 = \alpha(E = E_1)$ and $U_1 = U_d$.

The holes move towards the anode and create near it new electron-hole pairs after another period of time $\tau$. All these holes are collected at the anode and a new number of electrons $N_{02}$ start the second stage of the avalanche:

$$N_{02} = [\exp(\alpha_1 W) - 1] \times [\exp(\beta_1 W) - 1] \tag{3}$$

where $\beta_1 = \beta(E = E_1)$. And so, new electrons appear near the anode after every time period $2\tau$, starting a new stage of the avalanche.

There are two processes that affect electric field within the *i*-layer of the pixel. The first one is pixel discharge due to separation of electrons and holes and the second one is recharge of the pixel from an external power supply through the resistor $R_p$. Because of these processes, the second stage of the avalanche takes place at a different electric field strength $E = E_2 = U_2/W$ and different values of the ionisation coefficients $\alpha_2 = \alpha(E = E_2)$ and $\beta_2 = \beta(E = E_2)$. The new value of electric field is

$$E_2 = \frac{U_2}{W} = \frac{1}{W} \times \left[ U_1 - \frac{qN_1 - \frac{U_d - U_1}{R_p} \times 2\tau}{C_p} \right] \quad (4)$$

where $q$ is the electron charge and $C_p$ is the pixel capacitance. The value of the voltage drop due to the first stage of the avalanche is $qN_1/C_p$ and the value of the voltage increase due to recharge from the external power supply with voltage $U_d$ is $(U_d - U_1)2\tau/R_p C_p = 0$ for the first avalanche pass.

As a result, the number of electrons collected at the cathode after the second stage of avalanche process is

$$N_2 = N_{02} \times \exp(\alpha_2 W) = [\exp(\alpha_1 W) - 1] \times [\exp(\beta_1 W) - 1] \times \exp(\alpha_2 W) \quad (5)$$

Hence, the number of electrons collected at the cathode after the $i^{th}$ stage is

$$N_i = \left\{ \prod_{j=2}^{i} [\exp(\alpha_{j-1} W) - 1] \times [\exp(\beta_{j-1} W) - 1] \right\} \times \exp(\alpha_j W), \quad i \geq 2 \quad (6)$$

and the electric field at the same time is

$$E_i = \left(\frac{U_i}{W}\right) = \frac{1}{W} \times \left[ U_{i-1} - \frac{qN_{i-1} - \frac{U_d - U_{i-1}}{R_p} \times 2\tau}{C_p} \right]. \quad (7)$$

Gain (*M*) of the MAPD pixel for a single initial photo-electron can be calculated as

$$M = N_1 + \sum_{i=2}^{\infty} N_i. \quad (8)$$

Expressions (2) and (6) allow one to investigate dependence of the internal discharge current due to the avalanche process $I_i(t_i) = qN_{i-1}/2\tau$ and the external surcharge current $J_i(t_i) = (U_d - U_{i-1})/R_p$ at time $t_i = (i-1) \times 2\tau$. The time dependence of other parameters can be studied as well.

As it will be shown below, the proposed model can describe not only characteristics of MAPDs in Geiger mode (operating above the breakdown voltage) but also those of regular avalanche photo-diodes operating below the breakdown voltage.

## 3. Calculation and comparison with experimental data

One of the main MAPD parameters is its breakdown voltage $U_{br}$. It is defined as the minimum value of voltage applied between the cathode and anode of the pixel ($R_p = 0$) at which the avalanche process initiated by a single electron has infinite number of cycles. The latter condition is met when at least one electron is created at the beginning of each cycle, or

$$N_{0i} = N_{i-1}[\exp(\alpha_{i-1}W) - 1] \times [\exp(\beta_i W) - 1] = 1. \qquad (9)$$

Numerical solution of equation (9) using $W = 1$ µm, $v = 10^7$ cm/s, $C_p = 20$ fF, $R_p = 220$ kΩ and and $\tau = (W/v) = 10$ ps gives $U_{br} = 33.55$ V.

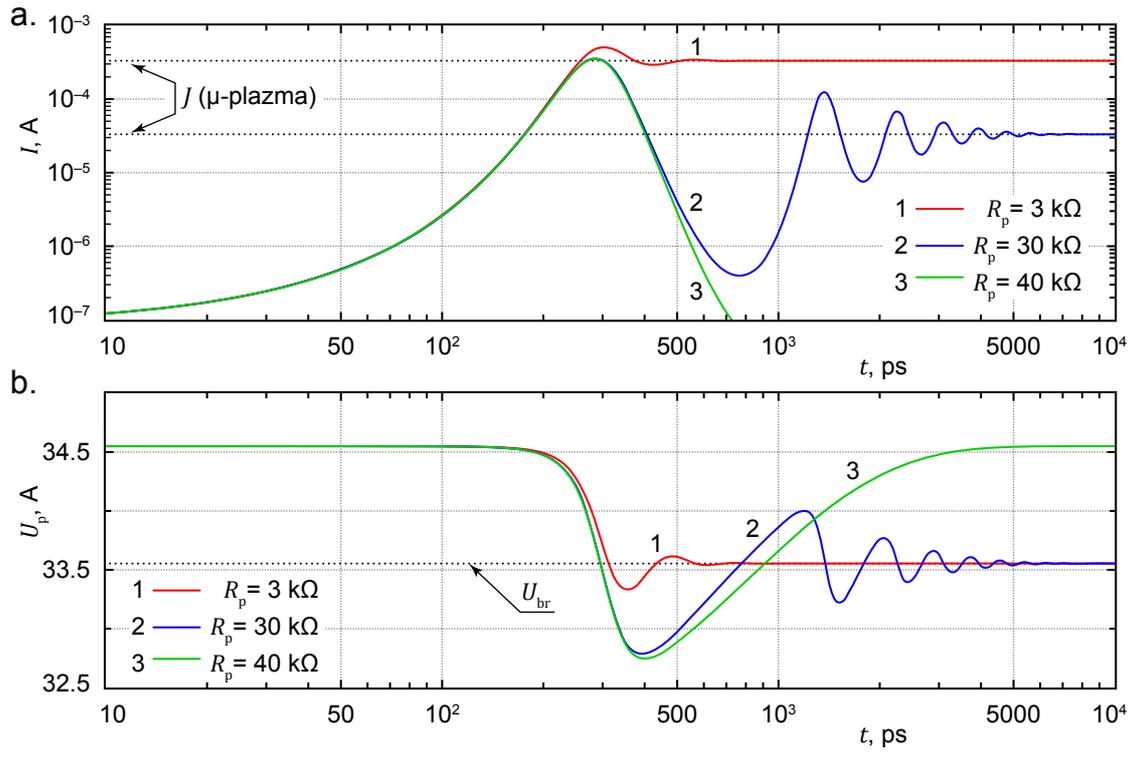

**Figure 2.** Temporal trace of the internal discharge current $I$ (*a*) and the pixel voltage $U_p$ (*b*) at values of the ballast resistor $R_p = 3$ kΩ (*1*), $R_p = 30$ kΩ (*2*), and $R_p = 40$ kΩ (*3*) and $C_p = \text{const} = 20$ fF.

Modelling results demonstrate that the mode of operation of an MAPD pixel depends both on the external voltage $U_d$ and on the pixel capacitance $C_p$. At certain fixed values of the internal pixel capacitance $C_p$ and over-voltage $\Delta U_p$, there is a corresponding (low) value of resistance $R_p$, at which the triggered avalanche becomes self-sustained, relatively quickly developing into a stationary current $I$, not unlike the already mentioned well-known micro-plasma breakdown in *p-n* transitions (see **Figure 2***a*, curve 1). At relatively large values of resistance $R_p$, a decaying transient oscillation of the current is observed (**Figure 2***a*, curve 2), also leading to the micro-plasma breakdown process, in which the conditions $U_p = U_{br}$ and $I = J$ hold true. In both these cases, the voltage on the pixel drops down to the breakdown value. This behaviour results from a high recharge current that does not let the pixel discharge below the breakdown voltage and quench the avalanche process. Even higher values of $R_p$ lead to termination of the avalanche process and establish Geiger mode of operation (**Figure 2***a* and 2*b*, curve 3). This indicates existence of a certain threshold value $R_{th}$ of the ballast resistor $R_p$, below which Geiger mode cannot be reached.

Physically, for the avalanche process to be quenched after the $i^{th}$ cycle (or at the moment $t_i = t_q$), it is necessary that the average number of electrons $N_{0i}$, which initiate the following avalanche cycle, be inferior to one, that is:

$$N_{0i} = N_{i-1}[\exp(\alpha_{i-1}W) - 1] \times [\exp(\beta_i W) - 1] < 1. \quad (10)$$

It was discovered that at a fixed breakdown voltage $U_{br}$, the introduced above threshold value $R_{th}$ depends on the bias voltage $U_d$ (or over-voltage) and the pixel capacitance $C_p$. As it can be seen from **Figure 3**, the value of $R_{th}$ reaches its maximum at bias voltage $U_d$ close to the breakdown voltage $U_{br}$ and then monotonically drops off at higher values of $U_d$. Physically, this non-trivial behaviour is caused by simultaneous discharging and charging processes, which depend differently on the over-voltage value. Of practical interest is the maximal value of $R_{th}$, which would guarantee that a pixel with capacitance of $C_p$ and breakdown voltage of $U_{br}$ may operate in Geiger mode at any bias voltage $U_d > U_{br}$. We will call this the critical value $R_{cr}$ of the ballast resistor. **Figure 3** presents three different values of the critical resistance $R_{cr}$ calculated for three values of the pixel capacitance $C_p = 20, 40, 80$ fF and equal to 220, 155, and 110 kΩ correspondingly.

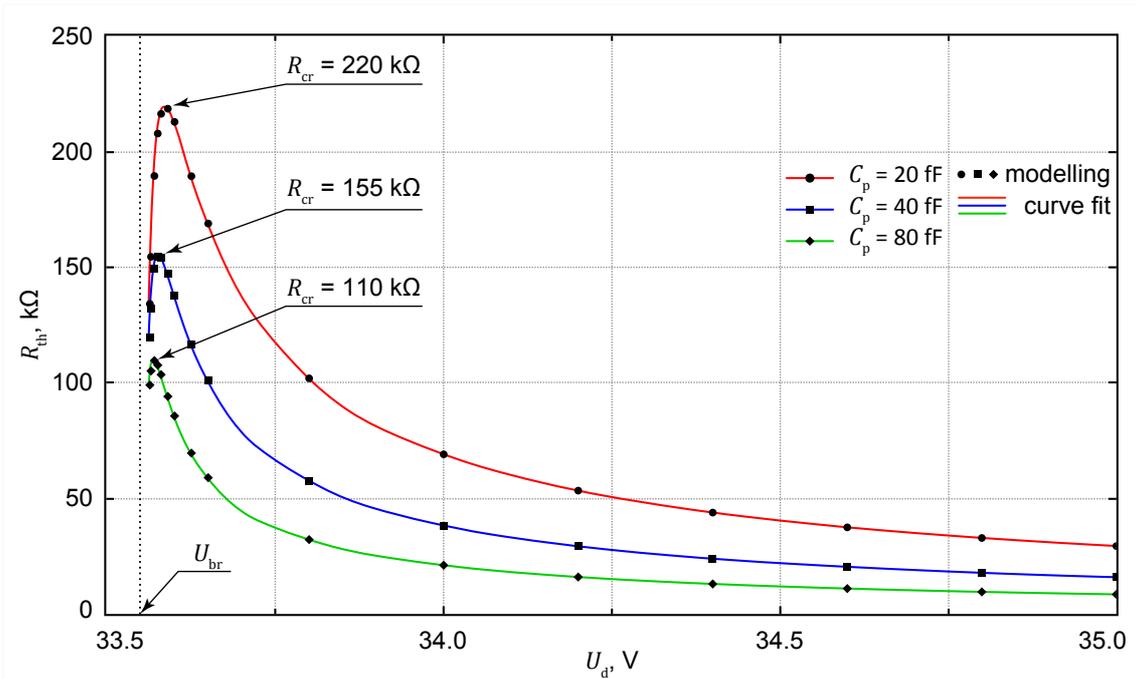

**Figure 3.** Dependence of the threshold resistance $R_{th}$ upon voltage $U_d$ applied to the pixel. Curves 1–3 correspond to capacitance $C_p$ values of 20, 40, and 80 fF respectively.

**Figure 4** shows time dependence of the internal discharge current $I$ (curve 1) and current in the external circuit $J$ (curve 2). Internal discharge current profile $I$ has the same values of both rise and fall times (about 44 ps at over-voltage $\Delta U_p = 2$ V). The same value of the rise time is shared by the external current $J$. However, the fall time of the external current pulse $J$ is three orders of magnitude longer. It is defined by the value of $C_p \times R_p \sim 4.4$ ns.

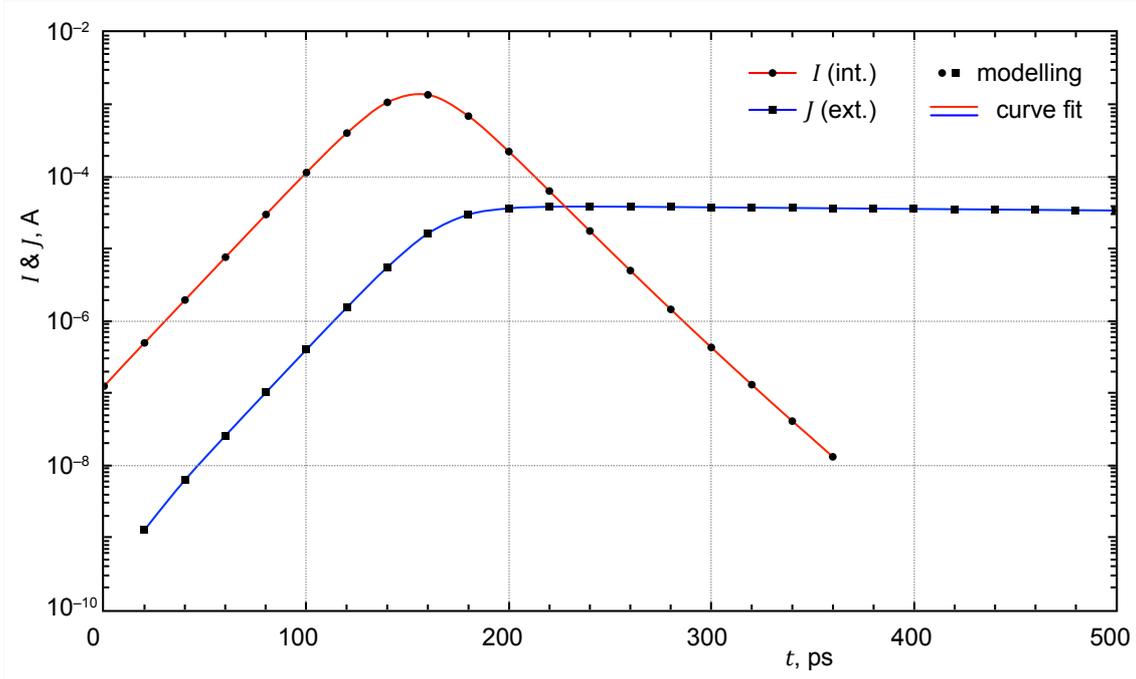

**Figure 4.** Shape of the MAPD pixel current pulse: internal avalanche current $I$ (red curve) and external current $J$ in the pixel circuit (blue curve).

In order to understand the internal process of avalanche development, it is important to know number of electrons $N_i$ created during each cycle of the avalanche process. Calculations show that the value of $N_i$ rises sharply within a few cycles and reaches $1.7 \times 10^5$ electrons. Thereafter, it falls at the same rate (**Figure 5***a*). During the same period, electric potential of the pixel $U_p$ decreases, reaching level of $U_{br}$ at the moment when the number of charge carriers is at its maximum. These charge carriers leads to further drop of pixel potential well below the breakdown voltage. As a result, the avalanche process is rapidly quenched. **Figure 5***b* shows that the total potential drop reaches $4\,V = 2 \times \Delta U_p$, where $\Delta U_p = (U_d - U_{br})$ is over-voltage. The apparent ~10-ps shift of the moment $U_p = U_{br}$ in **Figure 5***b* illustrates limited precision of the proposed model coming from the fact that the current value at stage $i$ is calculated from the parameters of the previous stage $(i-1)$ of the avalanche process. Physically, the maximum number of charge carriers in the avalanche cycle must be reached at $U_p = U_{br}$.

The total charge of a single-electron pulse ($Q_e = q \times M$) generated in an MAPD pixel is shown in **Figure 6** as a function of applied voltage $U_d$. It is common to believe that this linear dependence can be used to determine the pixel capacitance $C_p$ taken to be equal to $C_{eff} = \partial Q_e / \partial U_d$. Our model demonstrates that this is not the case and $C_{eff} = 2C_p$. As a result, $Q_e = C_{eff} \times \Delta U_p = C_p \times 2\Delta U_p$.

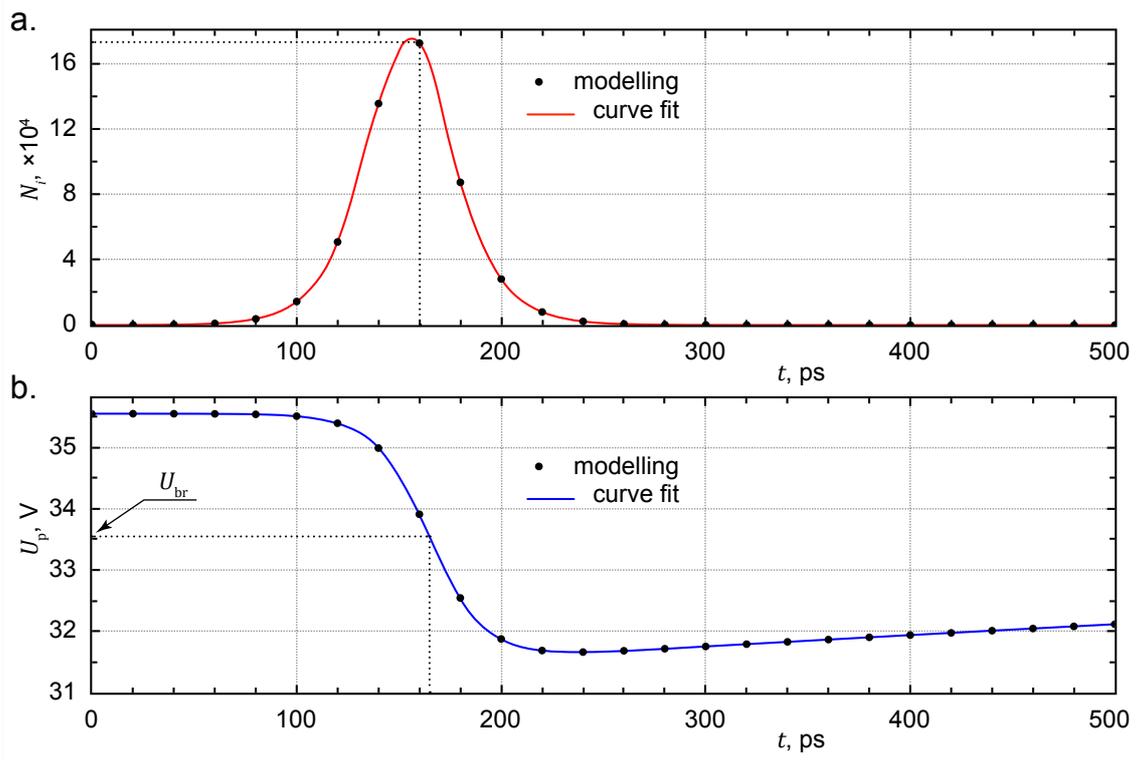

**Figure 5.** Time dependence of quantity of electrons $N_i$ generated during avalanche cycles (*a*) and voltage drop $U_p$ in the pixel (*b*).

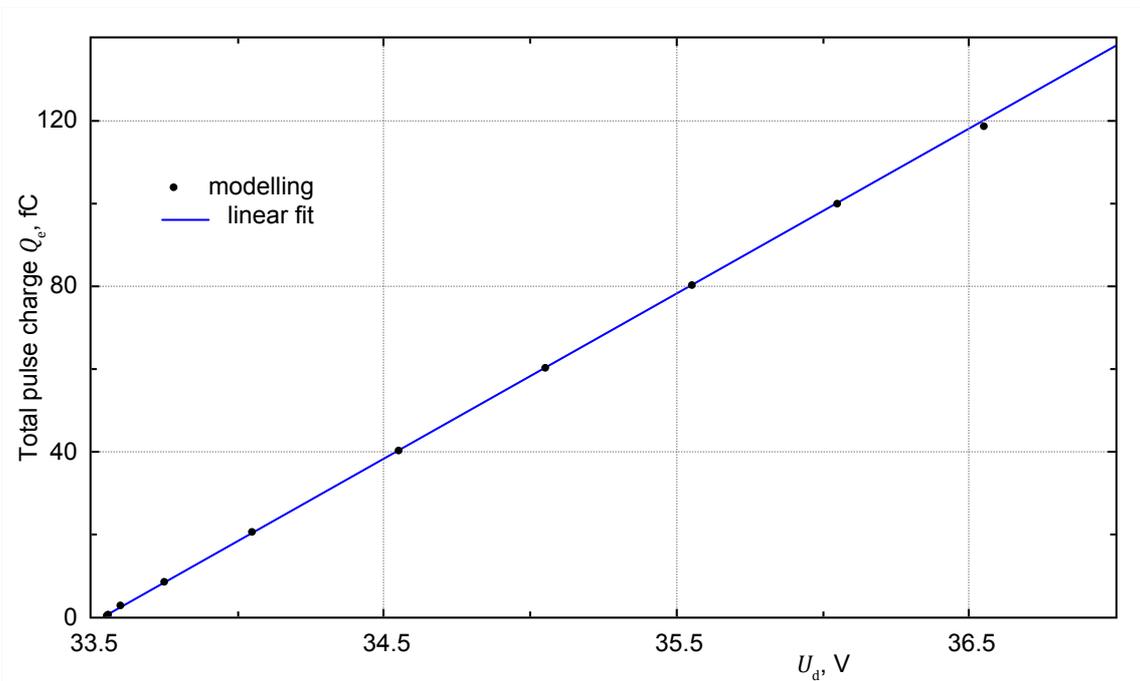

**Figure 6.** Dependence of total charge of a single-electron pulse on applied voltage ($C_p = 20$ fF, $R_p = 220$ kΩ).

Taking into account the preceding discussion of our results, it is possible to propose two new equivalent circuits of an avalanche photo-diode (pixel) operating in Geiger mode. The first equivalent circuit is an analogue of the original Haitz's model which comprises serially connected fictitious voltage source $U_0 = U_{\mathrm{br}} - \Delta U_{\mathrm{p}}$ (minimal pixel voltage), resistance $R_{\mathrm{s}}$ (resistance of space charge region of the photo-diode), and bi-stable switch $S$. The internal pixel capacitance $C_{\mathrm{p}}$ is connected in parallel to these components and the device is reverse-biased through ballast resistor $R_{\mathrm{p}}$ by voltage source $U_{\mathrm{d}} > U_{\mathrm{br}}$ (Figure 7a). This model is distinguished from the original one proposed in [8] by a different value of the internal voltage source (the original model has $U_0 = U_{\mathrm{br}}$).

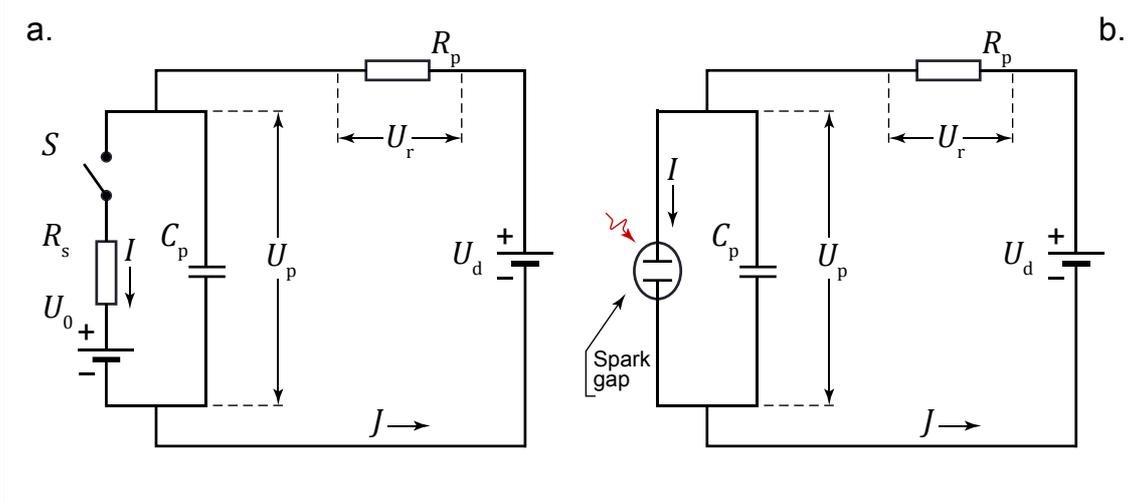

**Figure 7.** Equivalent circuits of a single MAPD pixel operating in Geiger mode.

The second equivalent circuit is based on our model discussed in the foregoing sections of this article. It is different from the previous one in that the avalanche photo-diode (pixel) is now represented by the internal device capacitance $C_{\mathrm{p}}$ and a spark gap with external quenching resistor $R_p$ (**Figure 7b**). The operation of this spark gap is governed by Equations (6)–(8).

In order to model various types of micro-pixel avalanche photo-diodes, the proposed circuits may be complemented with additional necessary elements (serial resistors, shunting capacitances, &c). However, within the scope of the present work, we limit our comparative analysis to the above-discussed circuits.

Applying the Kirchhoff's rules to the equivalent circuit of **Figure 7a**, we arrive at the equations for the dependence of the pixel voltage $U_{\mathrm{p}}$ and the external current $J$ upon time:

$$U_{\mathrm{d}} = U_{\mathrm{p}} + U_{\mathrm{r}}; \quad J = \frac{U_{\mathrm{r}}}{R_{\mathrm{p}}}; \quad I = \frac{U_{\mathrm{p}} - U_0}{R_{\mathrm{s}}}. \tag{11}$$

In the preceding Equation (11), $U_{\mathrm{r}}$ is the voltage drop over resistor $R_{\mathrm{p}}$, $U_0 = U_{\mathrm{d}} - \Delta U_{\mathrm{p}}$, and $\Delta U_{\mathrm{p}}$ is the over-voltage value. Additionally, voltage $U_{\mathrm{p}}$ may be expressed in terms of charge $Q_{\mathrm{p}}$ accumulated in capacitor $C_{\mathrm{p}}$:

$$U_{\mathrm{p}} = \frac{Q_{\mathrm{p}}}{C_{\mathrm{p}}} = \frac{1}{C_{\mathrm{p}}} \times (C_{\mathrm{p}} \times U_{\mathrm{d}} - \int_0^t I \times dt' + \int_0^t J \times dt'). \tag{12}$$

– 9 –

Differentiating Equation (12) and replacing $I$ and $J$ with their expressions from Equation (11), we will obtain:

$$\frac{dU_p}{dt} = -\frac{U_p}{\tau_r} + \frac{U_{st}}{\tau_r}; \quad U_p(t=0) = U_d, \tag{13}$$

where $U_{st} = \frac{U_0 \times R_p + U_d \times R_s}{R_p + R_s}$ is the stationary value of $U_p$ after switch $S$ makes the circuit (*i.e.* at $t \to \infty$), $\tau_r = R_o \times C_p$, and $R_o = \frac{R_p \times R_s}{R_p + R_s}$. By solving Equation (13), one can derive the following formulae for dependence of $U_p$ and $J$ upon time:

$$U_p(t) = U_{st} + (U_d - U_{st}) \cdot \exp\left(-\frac{t}{\tau_r}\right),$$
$$J(t) = \frac{U_d - U_{st}}{R_p} \times \left[1 - \exp\left(-\frac{t}{\tau_r}\right)\right]. \tag{14}$$

**Figure 8** demonstrates the results of calculations based on two equivalent circuits proposed in **Figure 7**. The value of $R_s$ was chosen so as to produce similar duration of the rise edges of photo-current pulses generated in the proposed circuits (around 40 ps). Here, we analyse the case when the avalanche process is triggered in both circuits simultaneously at $t = 20$ ps.

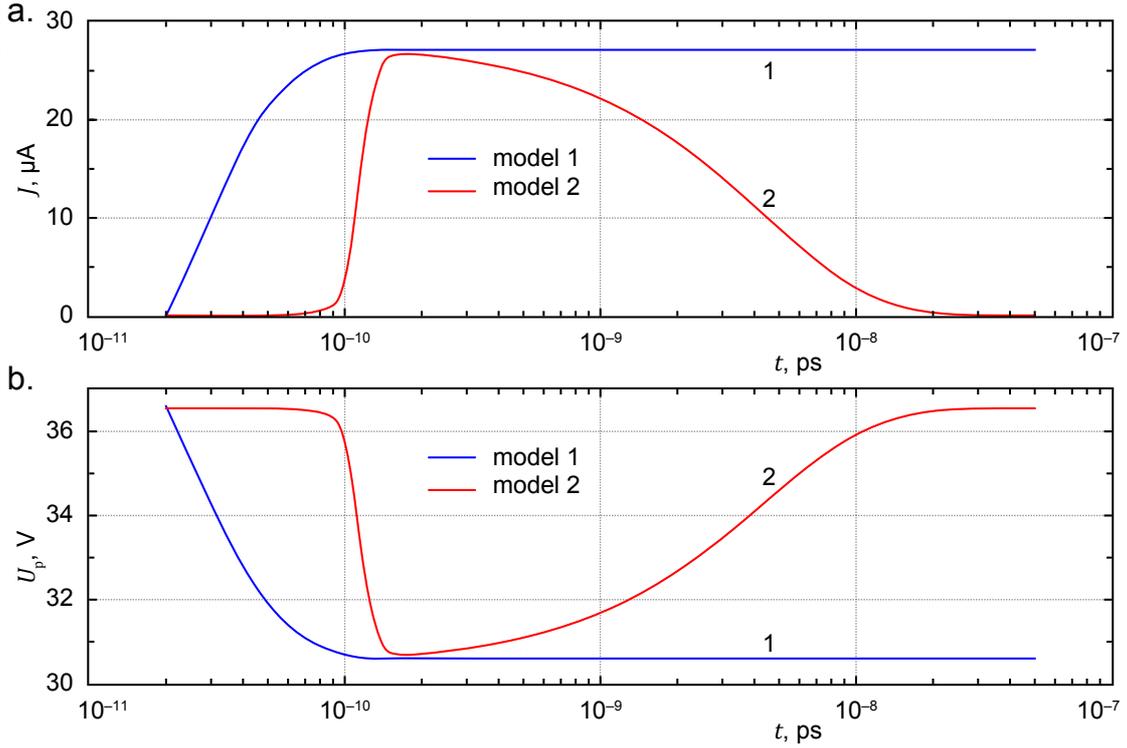

**Figure 8.** Temporal dependence of the external photo-current and pixel voltage calculated on the basis of the first (curve 1) and second (curve 2) equivalent circuits at $U_{br} = 33.55$ V, $\Delta U_p = 3$ V, $C_p = 20$ fF, $R_p = 220$ kΩ, and $R_s = 1$ kΩ.

It can be seen that the first of the discussed equivalent circuits is only adequate for description of the front edge of the photo-response, following which the external current reaches a certain stationary value. The pixel voltage drops below the breakdown voltage by the over-voltage value and remains at this level. Furthermore, the first of the proposed circuits does not produce a certain delay in formation of the photo-current pulse. This indicates that no equivalent circuit of this type can adequately describe the process of formation of a single-electron photo-current pulse in MAPD. The second equivalent circuit (**Figure 7***b*) obviously does not suffer from these limitations. This latter circuit demonstrates the entire process of formation of a single-electron avalanche photo-current with subsequent quenching of the avalanche process and restoration of the initial pixel voltage (curve 2 in **Figure 8**).

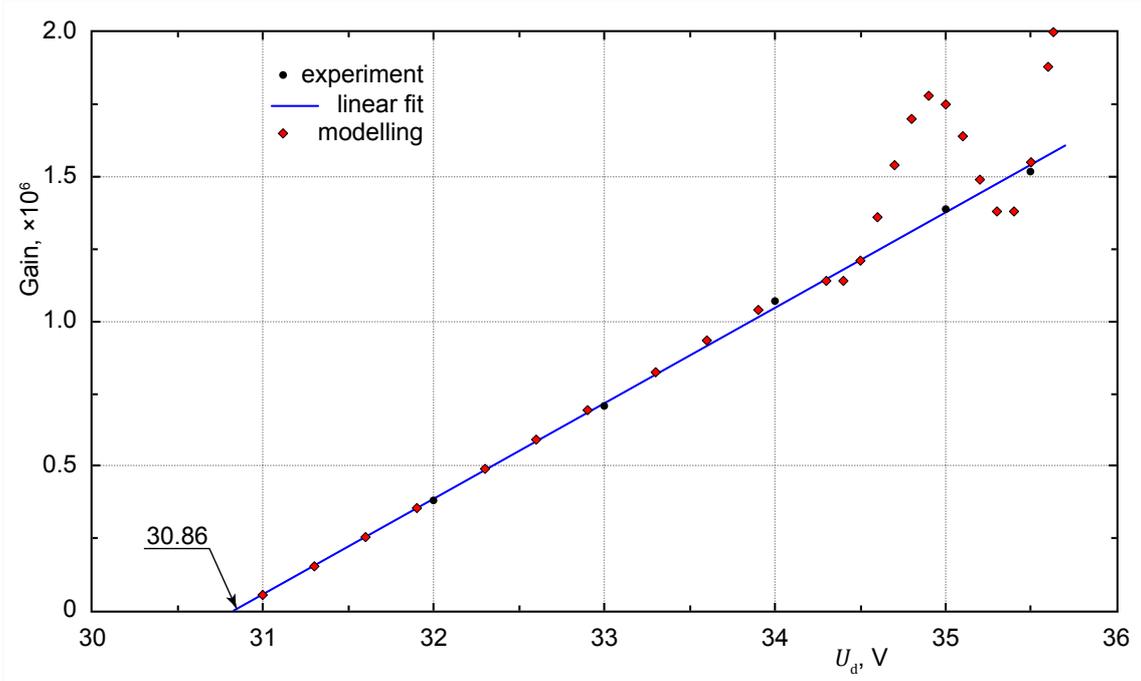

**Figure 9.** Gain of a single-pixel avalanche photo-diode as a function of the bias voltage.

In the following discussion, the results of our simulations are compared with some experimental data. It should be noted that our simple model does not take into account possible non-uniformity of electric field inside the experimental devices. However, various Geiger-mode devices demonstrate similar behaviour at the same over-voltage value $\Delta U_\mathrm{p}$. In our calculations, we selected model parameters so as to produce breakdown voltages close to those of the experimental devices. **Figure 9** shows an experimental gain-voltage dependence of a single-pixel avalanche photo-diode with an integrated individual micro-resistor (data taken from [22]). The value of the pixel's capacitance determined from the dependence slope is $C_\mathrm{eff} = 54$ fF. However, no experimentally measured data on pixel capacitance are given in [22]. In order to find the terminal capacitance of this device we used the effective pixel size (34×34 μm$^2$) and thickness of the depletion region (4 μm). Our calculation (in the approximation of a flat capacitor with no edge effects) based on these pixel dimensions taken from [22] gives a value of $C_\mathrm{p} = 30.4$ fF $\approx C_\mathrm{eff}/2$, which is predicted by the proposed physical model of avalanche process.

Results of modelling at $U_\mathrm{br} = 30.86$ V and $C_\mathrm{p} = 30.4$ fF are also presented in **Figure 9**. One can see that the proposed model describes the experimental data reasonably well up to $U_\mathrm{p} = 34$ V (equivalent to over-voltage value of $\Delta U_\mathrm{p} = 34 - 30.86 \approx 3$ V). Considerable discrepancy of modelling and the experimental data at $U_\text{p} > 34$ V arises from rapid increase of both electron and hole ionisation coefficients within the avalanche region. The pixel is discharged considerably in very few multiplication cycles, stopping the avalanche process. Rapidity of quenching is driven by the number of charge carriers produced in the last avalanche cycle, in which $U_\text{p} \geq U_\text{br} = 30.86$ V. Since in order to calculate the number of carriers in each cycle of the avalanche process ($N_i$), the data from the preceding cycle ($N_{i-1}$) are used, limited number of cycles leads to substantial deviation from the experimental data, including apparent oscillations in the dependence of gain upon bias.

We also studied our SPAD-type device fabricated together with Zecotek Photonics, Inc. (www.zecotek.com), which consisted of two elements: a small-area photo-diode and an external quenching resistor $R_\text{p} = 200$ kΩ. Both of these elements were mounted on a PCB (printed circuit board) plate. The small-area photo-diode was fabricated on the basis of a 3-µm thick epitaxial silicon layer with *n*-type conductivity grown on top of a silicon substrate having *n*-type conductivity. Specific resistance of the substrate and the epitaxial layer was 0.05 and 30 Ω·cm respectively. The active and contact areas of the photo-diode were 100×100 and 160×160 µm correspondingly. The contact area was wire-bonded to the quenching resistor $R_\text{p}$. Total terminal capacitance of this SPAD device was $C_\text{p} = 1.3$ pF.

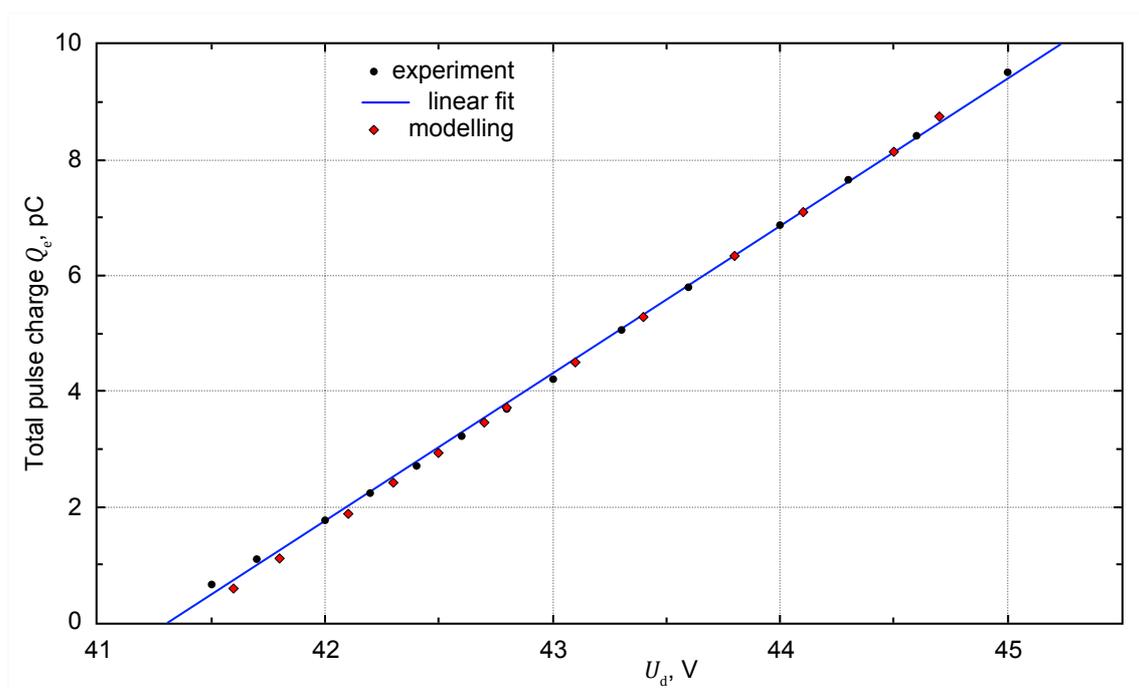

**Figure 10.** Charge of single-photoelectron pulses as a function of the bias voltage measured in the Zecotek device.

**Figure 10** shows dependence of the single-photoelectron charge upon the applied voltage. One can see that results of our modelling demonstrate good agreement with the experimental data. Calculated from the line slope, effective pixel's capacitance is $C_\text{eff} = 2.7$ pF, which is approximately twice the value of the measured pixel capacitance $C_\text{p} = 1.3$ pF.

## 4. Conclusion

The developed iterative model allows one to simulate operation of avalanche photo-diodes in Geiger mode. It gives correct qualitative behaviour of MAPD parameters as a function of applied voltage.

Simulations show that typical duration of the front and rear edges of the discharge current are of the same magnitude, which is shorter than 50 ps at over-voltage value $\Delta U_\mathrm{p} = 2$ V. The leading front of the external recharge current is also of the same magnitude.

One of the important results obtained from this model is a description of the avalanche process behaviour when the pixel potential reaches the value of the breakdown voltage. The pixel potential then continues dropping and new electron-hole pairs are being produced at a decreasing rate. The number of charge carriers produced after the voltage drops below the breakdown voltage is approximately the same as the number of those produced above the breakdown voltage. As a result, the potential on the pixel drops below the breakdown voltage by the over-voltage value $\Delta U_\mathrm{p}$, and effective pixel capacitance calculated as the slope of linear dependence of the pulse charge on bias voltage is about two times greater than its terminal capacitance.